\documentclass{llncs_Ibergrid2013}%

\usepackage{makeidx}  

\usepackage[dvips]{graphicx}

\usepackage{todonotes}

\begin{document}

\frontmatter          
\pagestyle{headings}  
\addtocmark{Analysis of Scientific Cloud Computing requirements} 
\mainmatter              
\title{Analysis of Scientific Cloud Computing requirements}
\titlerunning{Ibergrid}  
%
\author{{\'A}lvaro L{\'o}pez Garc{\'i}a\footnote{e-mail of corresponding author: \email{aloga@ifca.unican.es}},
        Enol Fern{\'a}ndez del Castillo
}
%
%
\tocauthor{{\'A}lvaro L{\'o}pez Garc{\'i}a  (Instituto de F{\'i}sica de Cantabria, CSIC - UC, Spain),,
Enol Fern{\'a}ndez del Castillo  (Instituto de F{\'i}sica de Cantabria, CSIC - UC, Spain),}
\institute{Advanced Computing and e-Science Group\\
Instituto de F{\'i}sica de Cantabria, CSIC - UC, Spain\\
\email{aloga@ifca.unican.es}, \\
}

\maketitle              

\begin{abstract}
While the requirements of enterprise and web applications have driven the development of Cloud
computing, some of its key features, such as customized environments and rapid elasticity,
could also benefit scientific applications. However, neither virtualization techniques nor
Cloud-like access to resources is common in scientific computing centers due to the negative
perception of the impact that virtualization techniques introduce.

In this paper we discuss the feasibility of the IaaS cloud model to satisfy some
of the computational science requirements and the main drawbacks that need to be
addressed by cloud resource providers so that the maximum benefit can be obtained from
a given cloud infrastructure.

\end{abstract}

\section{Introduction}
\label{sec:introduction}

Nowadays Cloud computing has achieved great success in the enterprise world but
it is still not common in the scientific computing field. Virtualization ---that
is a key component on the Cloud computing---, and
its associated performance degradation, has traditionally been considered as not
compatible with the computational science requirements. However, nowadays it is
accepted that virtualization introduces a low CPU overhead 
\cite{Barham2003,Ranadive2008} and that the penalty introduced in I/O can be
significantly reduced with techniques such as SR-IOV \cite{sriov} and
PCI-Passthrough \cite{pcipass} that provide near native performance
\cite{Dong2012} using modern hardware with specialized support for virtualization.

Moreover, virtualization also brings some benefits that overcome its performance drawback,
namely isolation and encapsulation. The isolation of VMs prevents influences from misbehaving VMs
to impact on other running VMs, while encapsulation of VMs gives the means to provide load balancing
and high-availability techniques. Virtualization also enables the consolidation of services by
providing support for a wider range of services with the same physical hardware, that leads to a
more efficient usage of the infrastructure and a reduction of maintenance costs.

This work complements some previous studies, such as Blanquer~et~al.~\cite{Blanquer},
Ramakrishnan~et~al.~\cite{Ramakrishnan2011} and Juve~et~al.~\cite{Juve2010}. This paper is
focused in the set of requirements ---for a resource provider and the cloud middlware--- that
a IaaS Cloud should provide for scientific usage, therefore there are some
higher level aspects (namely programming models, job-oriented execution models,
etc.) that are not covered by this paper. Also, it is worth noting that we are not
focusing on higher level Cloud service models (such as PaaS or SaaS).

In Section~\ref{sec:benefits} we give an outlook of the main benefits of using
a Cloud Computing model for scientific research. In Section~\ref{sec:applications} a set of pilot
use cases is described. From this preliminary group of applications,
in Section~\ref{sec:requirements} we have identified and established some requirements for a
Scientific Cloud Infrastructure.

\section{Cloud Computing benefits for scientific applications}
\label{sec:benefits}

Cloud Computing can be defined as ``a model for enabling ubiquitous, convenient,
on-demand network access to a shared pool of configurable computing resources (e.g., networks,
servers, storage, applications and services) that can be rapidly provisioned and released with
minimal management effort or service provider interaction.''~\cite{nist}. This model
allows many enterprise applications to scale and adapt to the usage peaks without
big investments in hardware with a \emph{pay-as-you-go} model.

On the other hand, Scientific Computing can be defined as the efficient usage of
computer processing in order to solve scientific problems. It can be considered
as the ``intersection of numeral mathematics, computer science and
modelling''~\cite{Karniadakis2003} and spans a broad spectrum of applications
and systems, such as High Performance Computing (HPC), High Throughput Computing
(HTC), Grid infrastructures, small and mid-sized computing clusters, volunteer
computing and even local desktops.

Many of the features of the Cloud Computing model are already present in current
scientific computing environments: academic researchers have used
shared clusters and supercomputers since long, and they are being accounted
for their usage in the same pay-per-use basis ---i.e. without a fixed fee--- 
based on their CPU-time and storage consumption. Moreover, Grid computing
makes possible the seamless access to worldwide-distributed
computing infrastructures composed by heterogeneous resources, spread
across different sites and administrative domains. However, the Cloud
computing model fills some gaps that are impossible or difficult to satisfy and
address with any the current computing models in place at scientific
datacenters. In the following sections we describe the major benefits that the cloud computing
model can bring to a scientific computing infrastructure.

\subsection{Customized environments}

One of the biggest differences between the Cloud model and any of the
other scientific computing models (HPC, HTC and Grids) is the execution
environment flexibility. While in the later ones the execution environment is
completely fixed by the infrastructure and/or resource providers (e.g.
the European Grid Infrastructure\footnote{See http://www.egi.eu for
details}, one of the majors grid infrastructures with $300+$ resource centers
providing $320,000+$ cores, supports only 3 Operating System flavors), in the
Cloud model the execution environments are easily adaptable or even provided by
the final users. This makes possible the deployment of completely customized
environments that perfectly fit the requirements of the final scientist's
applications.

This lack of flexibility in the current computing infrastructures ---where a
specific (or a very limited group) operating system flavor with a specific set
of software and libraries is deployed across all the available computing
nodes--- forces most applications to go through a preparatory phase before being
executed to adapt them to the execution environment idiosyncrasies, such as
library and compiler versions.
Moreover, some scientific applications use legacy libraries that are not compatible
with the available environments, rendering this preparation step quite
time-consuming or even impossible in some cases. The users could get rid of this
procedure to an extent if they were able to provide its own computing
environment, that will be the one used for its computations.

The requirement of a fixed operating system and the absence of customization
has been identified~\cite{Hoffa2008} as one of the main show-stoppers for
many scientific communities to adopt Grid computing technologies. Only large
communities are able to tackle this issue, thanks to dedicated manpower to
manage and adapt their software development and deployment to the available
scenarios.

Providing custom execution environments independently of the underlying physical
infrastructure also allows long-term preservation of the application environment
and opens the possibility of running legacy software with current and future
hardware, which may help in the long-term preservation of data (and analysis
methods for those data) of scientific experiments.

\subsection{On-demand access with rapid elasticity}

The Cloud model is based on on-demand and pay-as-you-go access that gives
the illusion of infinite resource capacity that can rapidly adapt to the needs
of the user. Although providing an infinite resource capacity is not feasible in
scientific datacenters, on-demand access to resources is useful for interactive
tasks.

Resources in the cloud model are elastically provisioned and released, opening
the door to using disposable environments without the overhead of a physical
deployment would imply (hardware preparation, re-installation, configuration).
These kind of disposable environments can be used for large-scale
scalability tests of parallel applications, or for testing new code or library
versions without disrupting production services already in place.

\subsection{Non-conventional application models}

Most scientific computing resources (supercomputers, shared clusters and grids)
are focused on processing and execution of atomic tasks, where each
of these tasks may be parallel or sequential and they may have interdependencies
between them or be executed concurrently. All the tasks have a common
life-cycle: they are started, they process some data and eventually return a
result.

However, in an Infrastructure as a Service (IaaS) Cloud, this traditional task
concept does not exist: instead of tasks, users manage instances of virtual
machines, which are started, stopped, paused and terminated according to the
needs of the users. This different life-cycle makes possible to create creation
of complex and dynamical long-running systems. For example this feature is used
in the simulation of dynamic software agents, as in~\cite{Sethia2011,Talia}; the
decision making process in urban management~\cite{Khan2012} or behavioral
simulations using shared-nothing Map-reduce techniques \cite{Wang2010}.

\section{Application use cases}
\label{sec:applications}

In this section we present some preliminary use cases deployed in our private
Cloud testbed. Although the applications are executed successfully in the
current infrastructure, we have identified some drawbacks that should be
addressed so that the scientific users could get even a better experience.
These topics will be further discussed and described on
Section~\ref{sec:requirements}.

\subsection{PROOF}

The Parallel Root Facility (PROOF)~\cite{Antcheva2009} is a commonly used tool
by the High Energy Physics (HEP) community to perform interactive analysis of
large datasets produced by the current HEP experiments. PROOF performs a
parallel execution of the analysis code by distributing the work load (input
data to process) to a set of execution hosts in a single program, multiple data
(SPMD) fashion.

PROOF is used in the last phases of the physics analysis to produce the final
plots and numbers, where the possibility of interactively change the analysis
parameters to steer the intermediate results facilitates the researchers work
and allows them to reach faster to better results. Data analyzed in this phase
contains the relevant physics objects in set of files ---produced by
several previous processing and filtering steps of the original raw data
collected from the detector--- that may range from several GBytes to a few
TBytes.

These analysis tasks are usually I/O bounded~\cite{Rodriguez-Marrero2012} due
to the big volume of data to process and their relatively low CPU
requirements: most codes perform filtering of the data according to the
relevant physics to be measured.

Running PROOF requires the pre-deployment and configuration of a master, that
acts as entry point and distributes the workload, and a set of workers where
the user's analysis code is executed. There are tools that automatize the 
creation of such deployments, which is not trivial for most users,
in batch-system environments~\cite{Rodriguez-Marrero2012,malzacher2010}, but
the machines are shared with other jobs, which may cause a degradation of the
performance.

A IaaS Cloud testbed provides support to these kind of interactive analysis
(i.e short lived sessions initiated on-demand by the users and with high
performance access to data) with customized environments where the
PROOF daemons run isolated from other workloads and are disposed as the
analysis finishes.

\subsection{Particle Physics Phenomenology}

As many other communities, the particle physics phenomenology groups develop
their own software for producing their scientific results. Software packages
developed by the community have evolved independently for several years, each of
them with particular compiler and library dependencies. These software
packages are usually combined into complex workflows, where each step requires
input from previous codes execution, thus the installation and
configuration of several software packages are mandatory to produce
the scientific results. Moreover, each scientific scenario to be analyzed may
require different versions of the software packages, therefore the researchers
need to take into account the different package versions characteristics for
installing and using them. Some of these packages also require access to proprietary
software (e.g. Mathematica) that is license-restricted. Although institutional
licenses may be available, these are difficult to control in shared resources (like grids or clusters) due to the
lack of fine grained access control to resources. 

Setting up a proper computing environment becomes a overhead for the everyday
work of researchers: they must solve the potential conflicts that appear when
installing them on the same machine; and the fixed execution supported by the
resource providers forces them to deploy the tools in ad-hoc clusters or even
their own desktops.

A cloud computing testbed allows these researchers to deploy a stable
infrastructure built with the exact requirements for their analysis where
each machine is adapted to the different scientific scenarios to be
evaluated, i.e. with the specific software versions needed for the analysis. The cloud
infrastructure should be able to enforce any usage or license restrictions for
proprietary software.

The possibility of creating snapshots of the machines also allows the
recovery of previous experiments easily without recreating the whole software
setup. These users would benefit from contextualization tools that automatically
sets up and handles any dependencies of the software packages needed for the
analysis upon machine creation~\cite{context}.

\subsection{Pattern Recognition from GIS}

The Vegetation Indexes (NDVI\footnote{Normalized Difference Vegetation Index}
and EVI\footnote{Enhanced Vegetation Index}) estimate the
quantity, quality and development of the vegetation in a given area \cite{vegetation}
by means of remote sensor data, such as satellite images. Using pattern recognition
techniques it is possible to analyze the behavior of this index, so as to make a
non-supervised vegetation classification. The analysis of such data also opens the
door to other applications such as fire detection, deforestation and vegetation
regeneration. For these data to be analyzed several specialized tools need to be
deployed, such as Modis (satellite data analysis tools), GRASS and GDAL (geospatial
libraries and tools), PROJ4 (cartographic data management) and R statistical
programming environment (along with a large set of additional R modules for
interacting with the other pieces of software).

These analysis were carried out in advance in the Grid so all the required
software had to be installed beforehand. This required from the intervention of
a local support team, so as to ensure the correct deployment of the tools and
applications. During this process, incompatibilities were found between the
dependencies of the required software and the operating system libraries
installed. This process delayed the start-up of the actual data process several
weeks. Moreover, the users faced a new computing environment and had
to be instructed on how to interact with the installed software in order to use
the correct versions that had to be installed in non-standard locations.
Finally, the data produced were stored in an external database ---that had also
to be deployed--- so that they could be finally accessed and analyzed by the scientists.

This use-case could profit from the Cloud computing testbed in two ways.
Firstly, they could deploy a ready to compute self-contained image, bundling all
the required software into it; and secondly, they could deploy their own 
infrastructure to store and retrieve their data. By doing so, they would
reduce the time needed to start with the analysis (as the software is ready to
be executed), the usage entry barrier (as they are deploying its own environment
and they are familiar with it) and leverage the management of the external
database service to the Cloud middleware (so they do not need to host a physical
server for it).

\section{Requirements for a Scientific Cloud Infrastructure}
\label{sec:requirements}

From our experience supporting the execution of the applications described on
Section~\ref{sec:applications} we have gathered some requirements that a
scientific cloud infrastructure should provide to its scientific users
(however, this is not an exhaustive list and they are not formal requirements).
We have classified these requirements in three groups: application level requirements,
for requirements relevant for easing the usage of cloud resources; specialized hardware, for
requirements related to high-performance access to specialized facilities; and enhanced scheduling
policies, for those policies that the cloud provider should adopt to provide an adequate service for
scientific users.

\subsection{Application level requirements}
\label{sec:req:app}

The deployment of customized environments is one of the biggest advantages of
the Cloud Computing model against any other \emph{traditional} paradigms, but it may also
represent a drawback for users that are not familiar with systems
administration. In this context, scientific application catalogs and 
contextualization mechanisms are needed.

\subsubsection{Scientific Application Catalogs}
\label{sec:scicatalog}

The Cloud Computing flexibility to deploy customized virtual machines has associated
the responsibility of create and manage them. Most scientific users are not
prone to create, manage and maintain their own system images, nor have the skills or
knowledge to perform those tasks in a secure and efficient way. These users may
profit from a Cloud infrastructure where a predefined set of supported images is
already deployed, containing a wide range of the software they need. This ready
to use Scientific Application Catalog can lower the entry curve for this new
infrastructure.

Another aspect of this application catalogs is the access to licensed and
institutional software ---that is, software specially designed and/or tuned to
be executed and integrated within an institution. In this cases, only restricted
access to the images will be provided to the users, so that only the allowed
ones are able to run the requested software. For example, access to shared
and clustering filesystems ---such as IBM GPFS, Lustre, etc--- can only be
provided to machines that are trusted and properly configured. Offering these
images in the catalog with restricted access, will give access to these
resources easily.

\subsubsection{Image contextualization}

The contextualization of images can be defined as the process of installing,
configuring and preparing software upon boot time on a pre-defined virtual
machine image. This way, the pre-defined images can be stored as generic and
small as possible, since all the customizations will take place on boot time.

The image contextualization is tightly coupled with the Scientific Application
Catalogs described in Section~\ref{sec:scicatalog}. The catalogs are useful for
bundling self-contained and ready to use images, but sometimes this is
something not feasible, because the required software evolves and changes
frequently its version, because the software is under a
development and debugging process and it is not practical to bundle it inside a
self-contained image or simply because it needs some user-defined data so that
it can be properly customized.

In those cases, instead of creating and uploading a new image for each
application version and/or modification (a tedious process that is a time
consuming task for the image creator), the installation and/or customization
can be delayed until the machine boot time. By means of this mechanism the newest
version can be automatically fetched and configured, or the defined and variable
user-data provided to the image. This is done by means of ready to use and 
compatible image that contains all the necessary dependencies and requisites
for the scientific applications to be installed. This contextualization-aware
images will then be launched with some metadata associated,
indicating which the software to install and configure.

Nowadays powerful configuration management tools exist that can help with the
implementation of the described contextualization mechanisms. Tools such as
\emph{Puppet} \cite{web:puppet}, \emph{CFengine} \cite{web:cfengine},
\emph{Chef} \cite{web:chef}, etc. make possible to define a machine
\emph{profile} that will be then applied to a machine, so that an given
machine will fit into that profile after applying it. However, these languages
and tools introduce a steep learning curve, so that the cloud middleware
should provide a method to expose the defined profiles to the users easily.

\subsection{Specialized hardware}
\label{sec:req:hardware}

Scientific Computing sometimes requires access to specialized hardware that
is not often present at a commercial provider, that is not focused towards
scientific computing.

\subsubsection{High performance communications}

Most parallel applications need low-latency, high speed network interconnects
(such as Infiniband or 10GbE) in order to be efficiently executed. These
interconnects are common in HPC environments, but they are not so common in
cloud providers. Moreover, this hardware normally does not have the support 
for being virtualized or shared between several virtual machines. In order to
give access to these devices two solutions exist: PCI passtrough with IOMMU or
Single Root I/O Virtualization (SR-IOV).

\subsubsection{High performance data access}

It is common that data oriented workloads demand high speed access towards the
data to be analyzed. In a cloud framework, the data is normally decoupled from
the instance that is running, meaning that it is being stored elsewhere not known
a priory by the user. For example, block devices can be attached from a central
storage location over the network (by means of Ata over Ethernet or iSCSI) to a
running instance. If access to the data is not efficient enough, the
computation will be executed on the node will suffer from a performance
penalty that will make it unusable.

\subsection{Enhanced scheduling policies}
\label{sec:req:scheduling}

Scientific applications need of enhanced scheduling policies that take into
account not only the requested and available resources, but also the kind of
execution that is going to be done and any special requirement that the 
scientific user may have.

\subsubsection{Instance co-allocation}

Some workloads require of the parallel execution of tasks across several nodes.
In this context, large requests have to be discriminated between non-dependent
and tightly-coupled or parallel nodes. Although the former can be provided in a
first-come, first-server basis; the later ones need of some advances scheduling
features, so that the collocation of instances makes possible that the user's
tasks can be properly orchestrated. This way, not only the resources should be
reserved in advance, but also the overheads and delays introduced by the cloud
management software \ref{sec:startup} should be taken into account so that the
instances have the same boot time.

\subsubsection{Short startup overhead}
\label{sec:startup}

When a request is made, the virtual images have to be distributed from the
catalog to the compute nodes that will host the virtual machines. If the catalog
repository is not shared or the image is not already cached by the compute
nodes, this distribution will introduce a penalty on the start time of the
requested nodes. This distribution penalty can be quite significant in large
systems, or when bigger requests are made by a user.

Large requests are common in scientific workbenches, so a mechanism should be
provided to ensure that these request are not penalized by this transfer time.
Some possible solutions could be to share the image catalog, to pre-schedule
the image transfers in advance or to utilize some efficient and intelligent
distribution methods for the requested images instead of downloading them from
a central location.

\subsubsection{Performance aware placement}

A virtual machine can potentially share the same physical host with other machines.
This can introduce a performance degradation if the machines are competing for the
utilization of the system resources. For example, two machines that are
executing some I/O consuming task can interfere between them. This issue has to be
handled by the scheduler so that two competing machines are not scheduled in the
same node.

On the other hand, as we already explained in Section~\ref{sec:req:hardware}, 
the requested nodes may need access to specialized hardware such as low-latency
interconnects (for example Infiniband), GPGPUS, etc. In these cases, not only the
scheduler has to be aware of these available resources, but also the cloud
middleware should be able to manage them. These resources are normally attached
to the virtual machines without being virtualized (that is, attaching the PCI
device directly to the node) so they deserve a different treatment.

\subsubsection{Spot and preemptable instances}

Long-running tasks are common in computational science. Those kind of workloads
do not require from interactivity and normally are not time-bounded. Such tasks
can be used to fill the computing infrastructure usage gaps, thus a better
utilization of the resources will be obtained. This is normally done in
traditional scientific infrastructures by means of several techniques, such as
backfilling, priority adjustments, task preemption and chekpointing.

However, a virtual machine can be transparently paused into a safe state that can be
resumed later on. This allows to create new execution types at a lower cost for
the users, such as the so called by some commercial providers
\emph{spot-instances}: machines that will run whenever
there is enough room for them, but that can be preempted, paused or even destroyed
by higher priority tasks. This is an interesting topic for the scheduling field of 
Computer Science: the usage of reverse-auction \cite{Roovers2012} and
other economical models \cite{Pueschel2009} opens the door to a better
utilization of the resources, by making attractive for some users to utilize the
infrastructure in the usage-valley periods whenever they can afford to pay the
price for those resources.

\subsubsection{Bare metal provisioning}

There may be some situations where the user needs to run a native operating
system instead of a virtual system. Some use-cases for the bare-metal provisioning
are the deployment of machines that need to access to a given hardware that
cannot be virtualized and/or directly attached to the virtual machine, non-x86
architectures, databases, etc.

\subsection{Absence of vendor lock-in}

Interoperability is an important feature for many communities. The usage of open
standards such as the Open Cloud Computing Interface (OCCI) \cite{web:occi},
Cloud Infrastructure Management Interface (CIMI) \cite{web:cimi} and Cloud Data
Management Interface (CDMI) \cite{web:cdmi} is way to avoid the vendor lock-in
that currently exists with many commercial cloud vendors. 

At a lower level it is also important also the adoption of the standards so as
to avoid the ``hypervisor lock-in''. In this context, the adoption of the Open
Virtualization Format (OVF) \cite{web:ovf} should be considered for the
distribution of virtual appliances.

\section{Conclusions}

Cloud computing has permeated into the IT industry in the last few years, and it
is nowadays emerging in scientific computing environments. However, there are
still some gaps that need to be filled so that the computational science could
completely benefit from it. In this paper we have made a sort review of the
advantages that a cloud computing model can offer to scientific users and how
either the middlewares and the resource providers need to adapt to satisfy
their new potential users.

Cloud middlewares are normally arising from the commercial providers ---being
Open Source or not--- and they are normally focused to satisfy their needs,
that are not the same as the scientific users requirements. From our
experience, one of the main fields needing from improvement in these middlewares
is the scheduling, as described in Section~\ref{sec:req:scheduling}. There is also
room for improvement in additional and higher level applications, such as
image catalogs and machine contextualization systems as described in
Section~\ref{sec:req:app}.

Scientific computing datacenters have to move
towards a mixed and combined model, where a given user can still access their
traditional computational power ---that is, using a batch system or using the
Grid---, but also they should provide their users with some additional cloud 
power that will complement the former computing models. This way, either the
users the users will benefit from a richer environment, and resource providers
can get a better utilization of their resources, since they will allow for new
execution models that are currently not available.

\bibliographystyle{ieeetr}
\bibliography{references}

\end{document}